\documentclass[a4paper]{article}

\usepackage{ISCSLP2021}
\usepackage{subfigure}
\usepackage{multirow}
\usepackage[colorlinks,linkcolor=black,anchorcolor=black,citecolor=black,urlcolor=black]{hyperref}
\title{ByteSing: A Chinese Singing Voice Synthesis System Using Duration Allocated Encoder-Decoder 	Acoustic Models and WaveRNN Vocoders}
\name{Yu Gu, Xiang Yin,   Yonghui Rao, Yuan Wan, Benlai Tang, \\ Yang Zhang, Jitong Chen, Yuxuan Wang,  Zejun Ma}

\address{  ByteDance AI Lab }
\email{ {\{guyu.ailab, yinxiang.stephen,  raoyonghui, wanyuan.0626, tangbenlai, \\ zhangyang.elfin,  chenjitong.1, wangyuxuan.11, mazejun\}@bytedance.com}}

\begin{document}

\maketitle
\begin{abstract}
This paper presents ByteSing, a Chinese singing voice synthesis (SVS) system based on duration allocated Tacotron-like acoustic models and WaveRNN neural vocoders.  Different from the conventional SVS models, the proposed ByteSing employs Tacotron-like encoder-decoder structures as the acoustic models, in which the  CBHG models and  recurrent neural networks (RNNs) are explored as encoders and  decoders respectively. Meanwhile an auxiliary phoneme duration prediction model is utilized to expand the input sequence, which can
enhance the model controllable capacity, model stability and tempo prediction accuracy. WaveRNN  vocoders are also adopted as neural vocoders to further improve the voice quality of synthesized songs. Both objective and subjective experimental results prove that the SVS method proposed in this paper can produce quite natural, expressive and high-fidelity songs by improving the  pitch and spectrogram prediction accuracy and the models using attention mechanism  can achieve best performance.
 
\end{abstract}

\section{Introduction}
\label{sec:intro}

Singing voice synthesis (SVS) systems can generate songs from the given musical scores which contain both linguistic information (lyrics) and different kinds of musical features such as note and tempo information. At present, SVS technique is an indispensable basic component 
in various applications with human-computer interaction such as virtual avatars, voice assistants and intelligent electronic devices. Meanwhile SVS systems can  be combined with
 other generation tasks such as  automatic lyric and melody generation. The assemble of multi-modal technologies, artificial intelligence singer and artificial intelligence composer has become more and more popular.
Thus the expectations for high-fidelity, high-naturalness and more accurate SVS algorithms will be increasing in future. 

Similar with text-to-speech (TTS) synthesis systems which only depend on linguistic input features, SVS systems generally adopt similar acoustic and duration models with statistical parametric speech synthesis (SPSS) systems. Some conventional statistical  models such as context-dependent hidden Markov models  \cite{saino2006hmm, oura2010recent} were employed on many popular SVS systems, which can model several acoustic features of a singing voice simultaneously. Nevertheless, suffering from the over-smoothing effect and limited modeling ability of those statistical models, the predicted musical features such as timbre and harmony by these systems had a great distinction with those extracted from ground-truth songs.

In recent years, deep learning technologies have achieved a resounding success on various speech and audio generation tasks \cite{dl_tts} such as TTS, voice conversion and speech enhancement.
Different kinds of models based on neural networks have also been proposed for SVS systems besides TTS systems. Deep neural networks (DNNs) and convolutional neural networks (CNNs) were employed to model the mapping relationship between the music scores and the acoustic features \cite{sinsy_dnn, nishimura2016singing,nakamura2019singing}. RNNs with long-short term memory (LSTM) cells were also adopted on SVS systems to capture the long range temporal dependencies  and produce higher singing quality \cite{kim2018korean}. 

Sequence-to-sequence models  such as Tacotrons \cite{wang2017tacotron, shen2018natural}  and  Deep Voice 3 \cite{deepvoice3} using the content-based attention mechanism are currently the predominant paradigm in end-to-end TTS, which have demonstrated the naturalness that could rival that of human speech. Some encoder-decoder structures for end-to-end  SVS \cite{angelini2019singing,lee2019adversarially} have also been proposed and adversarial training was adopted to improve accuracy of predicted features \cite{lee2019adversarially}.
 Despite these successes  of end-to-end models on TTS and SVS tasks, such methods usually suffer from a lack of robustness in the alignment procedure that leads to  repeated or skipped words  and incomplete
synthesis.  Different from the automatic and soft attention alignments in the conventional end-to-end models, some additional duration predictors were 
employed to address the issues of wrong
attention alignments and consequently reduce the ratio of the missing or repeating 
words \cite{ren2019fastspeech,yu2019durian}.
Similar duration informed attention network is also applied on SVS and singing conversion tasks \cite{ blaauw2019sequence,zhang2019learning} to ensure the hard alignments between the phoneme and musical score sequences and 
their corresponding acoustic features. Furthermore,  waveform modeling algorithms such as WaveNet \cite{wavenet},  WaveRNN \cite{wavernn} and WaveGlow \cite{prenger2019waveglow} have already achieved high-fidelity audio quality and  close-to-human perception and also been used in SVS systems \cite{yi2019singing}. 

Motivated by the achievements of acoustic models based on duration informed encoder-decoder architectures on audio generation tasks, we proposed a Chinese SVS system, ByteSing, to synthesize vocal waveforms from original musical scores  and lyrics following an end-to-end structure and an auxiliary phoneme duration prediction model. 
Different from the singing synthesis systems mentioned above \cite{blaauw2019sequence,zhang2019learning},  on which encoders were mainly dependent on the linguistic features and fundamental frequency (F0) trajectories, the proposed SVS system processes the  embeddings  of both linguistic and musical features. 
Besides, these systems were still dependent on traditional DSP vocoders and source-filter models, which
 were hard to extract accurate features from singing voice and had disadvantages of generating high quality waveforms.
 Therefore, the proposed ByteSing model utilizes an autoregressive decoder to convert the duration expanded input features into mel-spectrogram sequences directly, which contains more detailed and richer acoustic information. Phoneme duration information is also predicted to improve the model stability and the tempo  
 accuracy of synthesized songs. Furthermore,  WaveRNNs are adopted as vocoders to synthesize waveform directly to exceed the limitation of traditional vocoders. 
 
This paper is organized as follows. Section \ref{sec:method} introduces the detailed structure of the proposed system. In Section \ref{sec:experiments}, the performance of different system configurations is evaluated. The conclusion and future work are given in Section \ref{sec:conclusion}.
\section{The proposed system}
\label{sec:method}
\subsection{Overview}
\begin{figure}[t]
\setlength{\belowcaptionskip}{-0.2cm}
\centering
\includegraphics[width=\linewidth]{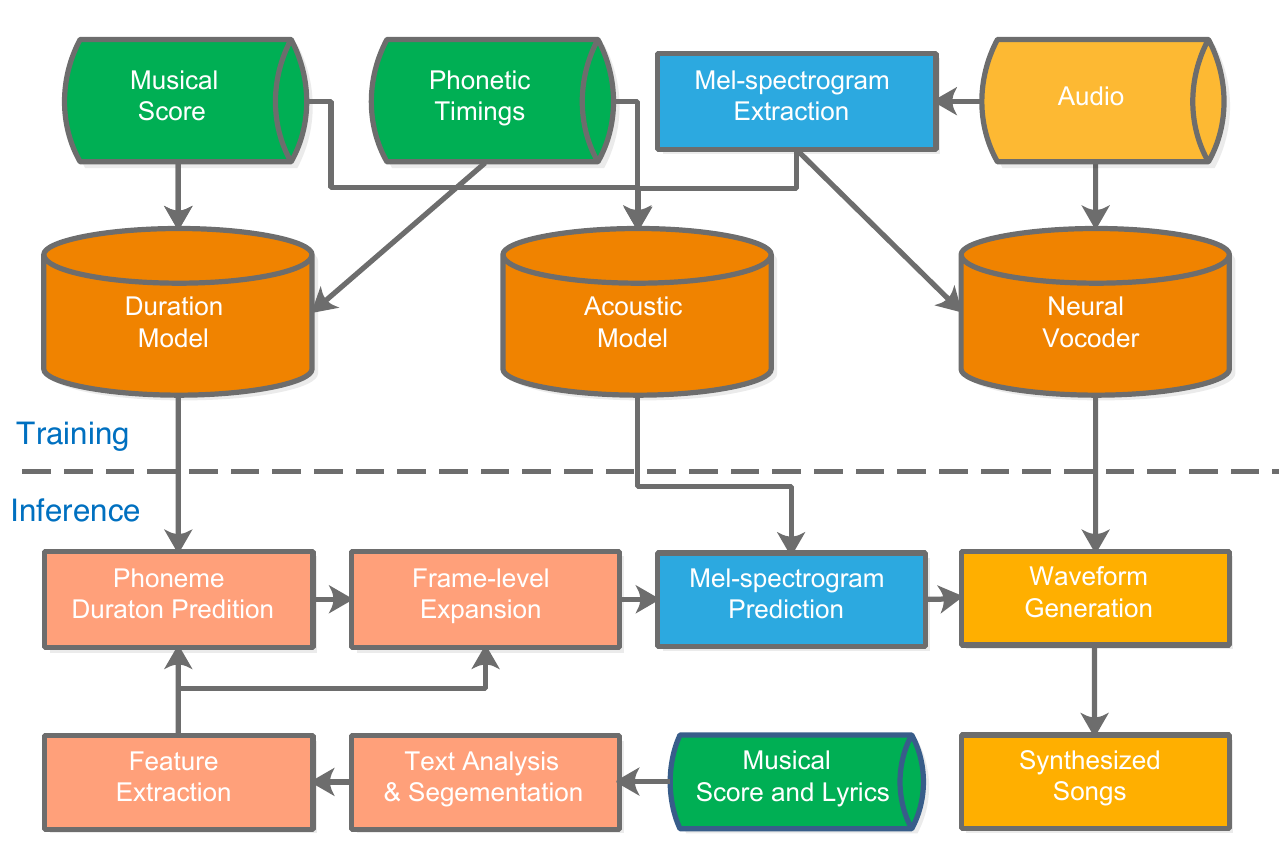}
\vspace{-0.3cm}
\caption{{Block diagram of ByteSing system.}}
\label{fig:flowchart}
\end{figure}
Figure \ref{fig:flowchart} depicts a general description of ByteSing system with its different
components.  To achieve the goal of imitating the timbre and the singing and pronunciation characters of the specific singer, a singing dataset is recorded following the given musical scores, which are described on  MusicXML format \cite{good2006musicxml}.  The recorded songs  are phonetically transcribed and segmented.
At the training stage,  a  duration model, an acoustic model  which is based on encoder-decoder framework and  a neural vocoder are trained respectively.  The duration model predicts the begin and end time of each phoneme using both linguistic and musical information and a post-processing step is conducted according to the  constraints of note durations. The note-level features are converted to frame-level  ones according to the interval information of adjacent phonemes.  The acoustic model is established to map the expanded frame-level input feature sequences into the extracted acoustic feature sequences. Different from other 
SVS systems which explicitly choose F0s and spectral envelope related features
 as acoustic features, 80-dimensional mel-spectrograms that  implicitly include all the acoustic elements such as pitch and format are directly predicted. 
 Meanwhile, a neural vocoder based on WaveRNN is constructed using recorded songs and  the mel-spectrograms extracted from  the corresponding ground-truth waveforms. 
 For the inference phase, some standard text analysis procedures such as 
  polyphone disambiguation are  performed on the score lyrics firstly to infer the 
  phoneme sequences for the lyrics. Long paragraphs are also segmented into short sentences for the convenience of modeling. Given the 
  phoneme sequences and their musical scores, the phoneme durations are predicted by the duration model. The frame-level expanded feature sequences are feeded into the encoder. Then the decoder generates the 
  mel-spectrogram sequence frame-by-frame  in an autoregressive manner.
  The trained neural vocoder can transform the predicted mel-spectrograms into singing waveforms. The details of each part on ByteSing system are described as follows.
  \subsection{Feature representation}
  \label{subsec:fr}
   \begin{table}[t]
\vspace{-0.2cm}
 \centering
 \renewcommand{\arraystretch}{1.1}
  \caption{ Definition and description of the symbols in this paper.}
  
 \vspace{-0.4mm}
   \begin{tabular}{ c c}
    \hline
    Symbol & Description (Examples) \\
     \hline \hline
      \emph{\textbf{Ph}}  & Chinese phoneme identities (sh, uai)   \\
      \emph{\textbf{Pi}}  & Pitch from note (C4, G3)   \\
        \emph{\textbf{Du}}  & Duration from tempo and note (0.625s) \\
        \emph{\textbf{Tp}} & Phoneme types (initial, final or zero-initial)   \\
        \emph{\textbf{To}} & Tone of the corresponding syllable (0, 1, 2, 3)\\
            \emph{\textbf{Po}} & Frame position embedding\\
            \hline
          \end{tabular} 
           \label{tab:symbol}
       \vspace{-0.2cm}
      \end{table}
   \vspace{-0.2cm}
\setlength{\abovecaptionskip}{0.2cm}
\setlength{\belowcaptionskip}{-0.2cm}
 We convert the musical scores and lyrics  into our  self-designed input feature sequences and 
 some self-defined symbols are described on the Table \ref{tab:symbol}.
  For the  duration models and acoustic models, two sets of feature composing are 
  adopted respectively. The input features $\bm{X}_{D}\!=\!\left[ \emph{\textbf{Ph}}, \emph{\textbf{Tp}}, \emph{\textbf{Du}}\right]$ in the duration models are phoneme-level ones where 
  $\emph{\textbf{Ph}}$  and $\emph{\textbf{Tp}}$ are both categorical  features and one-hot encoded and  $\emph{\textbf{Du}}$ is the theoretical numerical  duration of the note that current phoneme belongs to. $\emph{\textbf{Du}}$ can be obtained according to the tempo information and the note duration information.
   For the acoustic models, the duration expanded features  $\bm{X}_{A}\!=\!\left[ \emph{\textbf{Ph}}, \emph{\textbf{Pi}}, \emph{\textbf{Po}}\right]$ are frame-level ones where $ \emph{\textbf{Pi}}$ is also categorical ones rather than a 
  floating-point frequency value. $\emph{\textbf{Po}}$ is an additional three-dimensional  position embedding  computed as a ramp representing the advancements and reserve percentages of the phoneme for each frame  and position on the current utterance for each phoneme,  which are all normalised as  floating point numbers in the interval [0,1].
  \subsection{Duration models}
  Different from TTS tasks on which the duration is  only conditioned on the text contexts and  prosody characters of the specific speaker, the duration for singing should also refer to the musical duration which has less degrees of freedom than TTS.
 In the case of SVS,  the start timing and end timing of each 
   phoneme should be determined more accurately.  Therefore, 
   a bidirectional RNN incorporating multiple layers of LSTM cells is utilized  as the duration model to predict the duration of the target 
phoneme from the input phoneme-level $\bm{X}_{D}$ feature sequence. 
Then the supervised back-propagation through time algorithm  was
conducted to fine-tune the RNN parameters under minimum mean squared error (MMSE) criterion.

Although  the phenomenon  of time lags  between start timings of musical score and start timings of real audio  in actual songs is quite common, time-lag models  \cite{saino2006hmm} are not employed on ByteSing. On the contrary, for  procedure  simplification and the convenience of audio mixing with background music,  a post-processing step is performed on predicted phoneme durations to constrain the whole  syllable duration to be equal to the corresponding musical note durations. In fact, only the degree of freedom for the ratio of vowels and  consonants in each syllable is  reserved, which is a compromise of naturalness and veracity. Although the
musical scores are strictly followed which is not agreed with practice situation, the naturalness is not  degraded significantly from roughly subjective perception and the synthesized songs are more easily delivered to the post-productions such as automix and autotune.
\subsection{Acoustic models}
\begin{figure}[t]
\setlength{\belowcaptionskip}{-0.2cm}
\centering
\includegraphics[width=\linewidth]{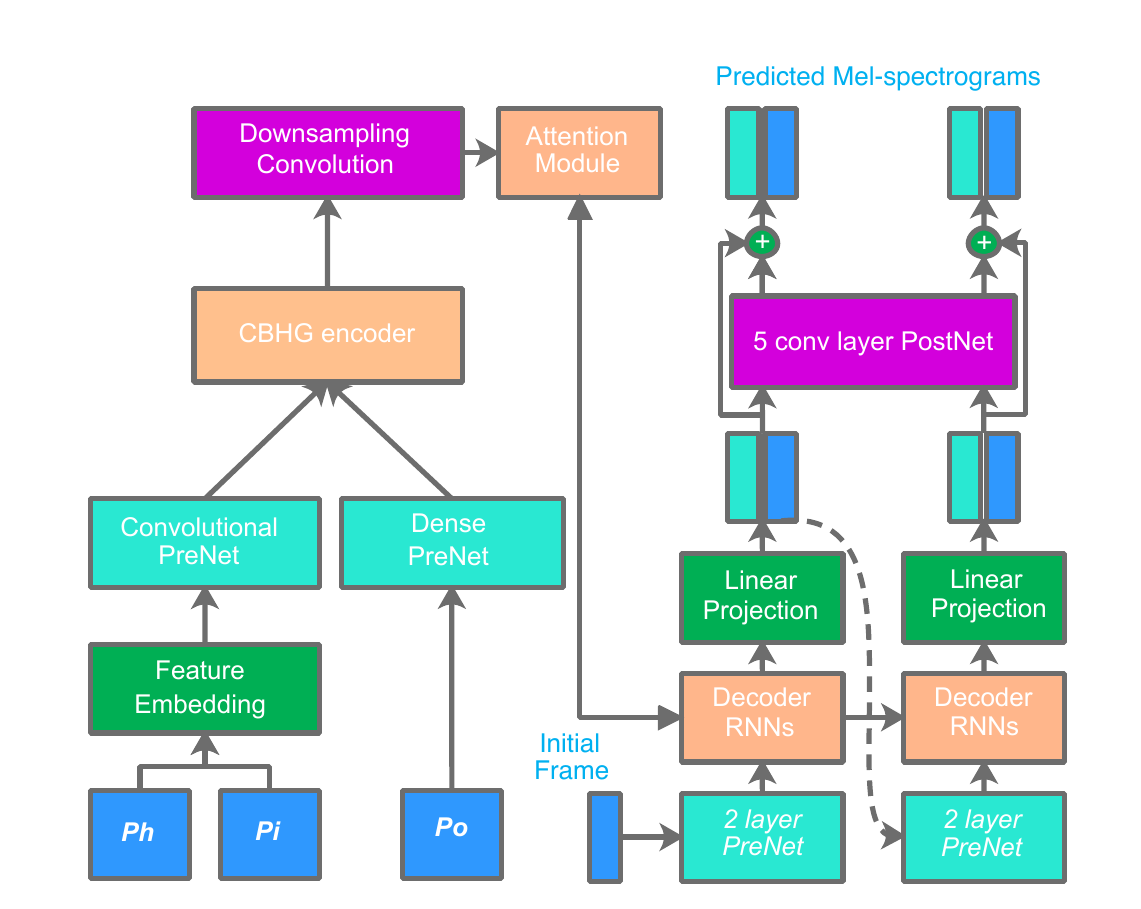}
\vspace{-0.3cm}
\caption{{Acoustic models in ByteSing.}}
\label{fig:acoustic}
\end{figure}
The  acoustic models used in ByteSing are depicted in Figure 
\ref{fig:acoustic}, which are totally evolved from both Tacotron and Tacotron2. The input sequence  $\bm{X}_{A}$ as presented on Section \ref{subsec:fr} is expanded to frame-level according to the  given phoneme durations. The 
$ \emph{\textbf{Ph}}$ and $ \emph{\textbf{Pi}}$ are categorical features and  encoded with embedding layers. A convolutional PreNet is deployed to model the  long-term information of both linguistic and musical  context. Inherited from Tacotron \cite{wang2017tacotron}, the powerful
CBHG module that consists of a
bank of  convolutional filters, highway networks and a
bidirectional gated recurrent unit based RNN is also used as the encoder for extracting representations from the input sequences. Then the encoded sequences are down-sampled to the  identical time-resolution with that of the output  acoustic feature sequences through a convolutional layer.
The  GMM-based attention mechanism \cite{graves2013generating,battenberg2019location} is exploited  to align the input musical and linguistic features with the output spectrograms. Because the input sequences are expanded by an auxiliary duration model, the attention module can achieve
 fast convergence and the monotonicity and locality properties of synthesis
alignment can also be guaranteed. Meanwhile, due to the attention strategy and encoder-decoder structure, the alignment between the source and target is learned automatically  and controlled by the dynamic context vectors, which has the advantages of  the hard alignments in conventional SPSS and SVS systems. The decoder of ByteSing follows the
decoder design of Tacotron2 \cite{shen2018natural}  and  an autoregressive RNN 
predicts mel-spectrograms from the encoded input sequence multiple
frames at a time. The acoustic prediction  from the previous time step is first
passed through a  pre-net containing 2 fully-connected layers.
 The output acoustic feature sequence from the decoder network is passed
through a  convolutional post-net to predict the residuals. Losses from before
and after the post-net are calculated to optimize the whole acoustic model.

\subsection{WaveRNN neural vocoder}
\begin{figure}[t]
\setlength{\belowcaptionskip}{-0.2cm}
\centering
\includegraphics[width=0.99\linewidth]{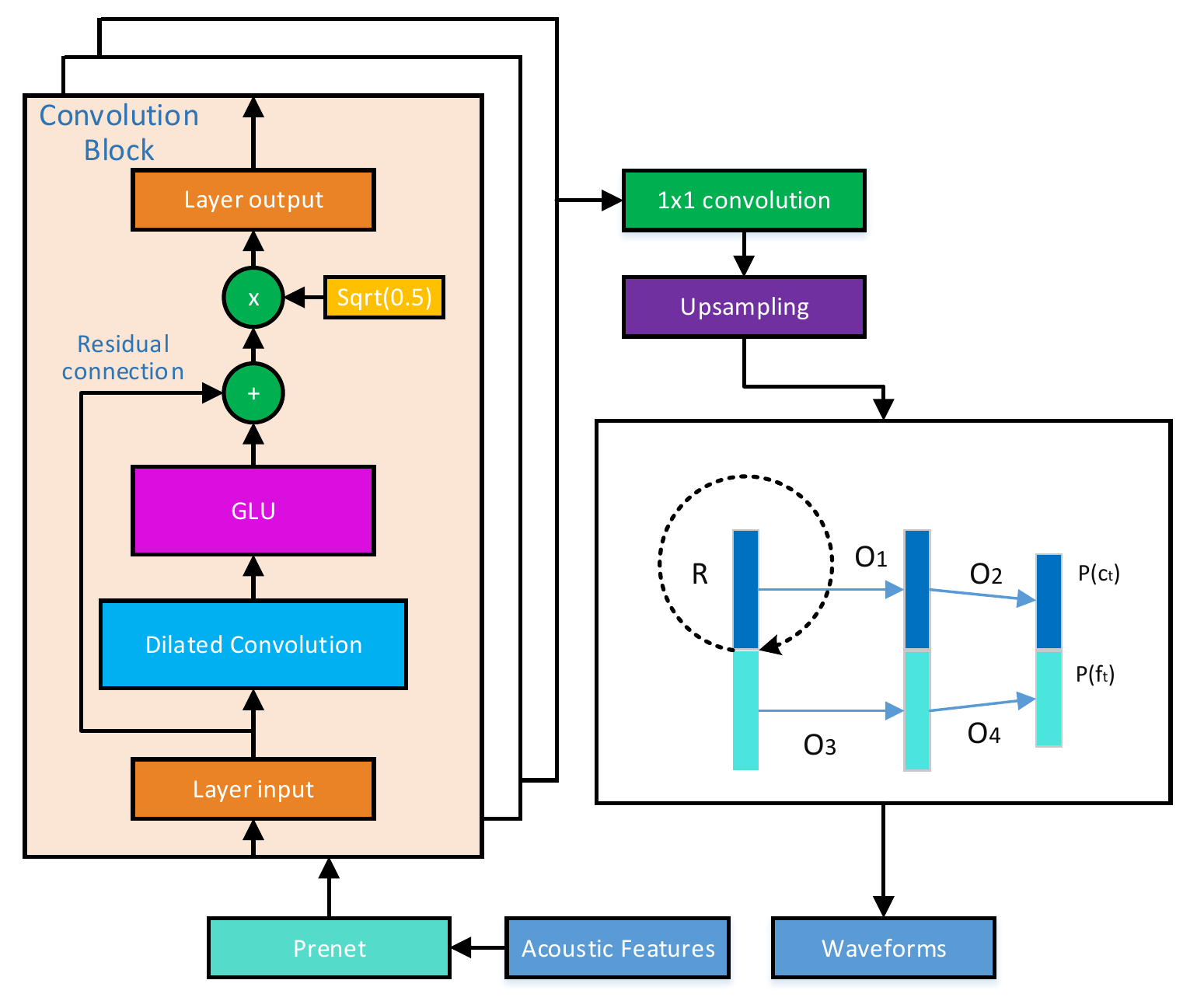}
\vspace{-0.1cm}
\caption{{Network structures of WaveRNN.}}
\label{fig:wavernn}
\end{figure}

WaveRNN \cite{wavernn} is a generative model which was proposed for TTS synthesis and other general audio generation tasks. WaveRNNs performed autoregressive speech sample generation  using  GRU variants instead of depending on vocoders, which predicted the coarse and fine part of audio samples successively. The architectures of  the exploited WaveRNN in ByteSing are illustrated on  Figure \ref{fig:wavernn}, which contains of  a sample generation network and  a condition  network. 
\label{sec:wavernn}
For the sample generation part, the original structure is basically 
followed using a single-layer RNN with a
dual softmax output layer to predict the categorical distributions of the audio samples conditioned on the predicted mel-spectrograms. Multiple layers of convolutional blocks as depicted in Figure \ref{fig:wavernn} are utilized to encode the frame-level
mel-spectrogram condition sequences, which is motivated by the  
encoder structure on Deep Voice 3 \cite{deepvoice3}.
The convolutional block consists of a 1-D convolutional layer,
a gated-linear unit (GLU)  \cite{glu} as a learnable nonlinearity, a residual connection to the input and a scaling factor of $\sqrt{0.5}$.
Stacked non-causal convolutional layers with an exponentially increasing the dilation factors can result in a sufficiently large receptive field and the GLU can  alleviate the vanishing gradient issue for stacked convolution blocks while retaining
non-linearity. The encoded information is upsampled to the same time resolution with native audio frequency by simply repetition and then added into the biases of GRU cells.

\section{Experiments}
\label{sec:experiments}
\subsection{Experimental conditions}
To evaluate the performance of the proposed ByteSing system, 90 Chinese songs 
performed by a female singer  were used as the training dataset. The recorded  songs
were finely  labelled  and were decomposed into short utterances according to the rest and  the lyric semantic information.  The sampling rate for the singing voice was  reduced to 24 kHz.
Another 10 songs those were not present  in the training set were used as the test dataset to measure the performances and the  musical scores and lyrics were annotated on MusicXML format. To evaluate the effects of attention mechanism and using different features,
the following several SVS systems were established for comparison.

\begin{itemize}
\setlength{\itemsep}{0pt}
\item \emph{\textbf{Natural}}:  The ground-truth recorded songs;
\item \emph{\textbf{ByteSing}}:  The proposed ByteSing system;
\item \emph{\textbf{BS-w/o-atten}}: ByteSing without attention module  and directly using predicted durations as alignments;
\item \emph{\textbf{BS-w-$\textbf{To}$}}: ByteSing also with tones (  \emph{\textbf{To}} in Table \ref{tab:symbol} ) as inputs to evaluate the effectiveness of tone information;
\end{itemize}

\begin{table}[t]
\vspace{-0.2cm}
 \centering
 \renewcommand{\arraystretch}{1.1}
  \caption{Comparison of distortion between acoustic features of natural voice and synthesized voice from different systems where  Corr. represents the correlation coefficients.}
  
 \vspace{-0.4mm}
        \setlength{\tabcolsep}{3mm}{
         \begin{tabular}{| c |ccc|}
            \hline
          System& \emph{\textbf{BS-w-$\textbf{To}$}} &   \emph{\textbf{ByteSing}}   &\emph{\textbf{BS-w/o-atten}}  \\
 
            \hline \hline
            \emph{\textbf{MSD}}  & 5.831  & \bf{ 5.763} &6.173 \\
 \emph{\textbf{F0 RMSE}}  & 32.930 & \bf{ 27.376} &33.896 \\
  \emph{\textbf{F0 Corr.}}  & 0.912 & \bf{ 0.940} &0.913 \\
            \hline
          \end{tabular}  
           }    
           \label{tab:obj}
      \end{table}
   \vspace{-0.2cm}
\setlength{\abovecaptionskip}{0.2cm}
\setlength{\belowcaptionskip}{-0.2cm}
 \vspace{-0.2cm}

\begin{figure}[t]
\setlength{\belowcaptionskip}{-0.2cm}
\centering
\includegraphics[width=0.7\linewidth]{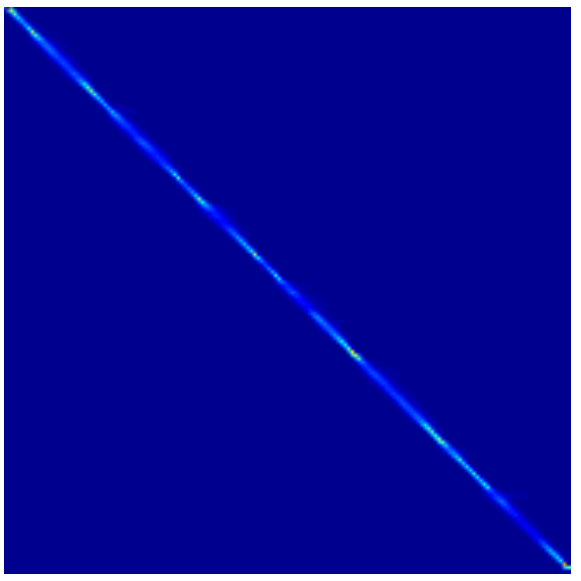}
\vspace{-0.1cm}
\caption{{Attention alignment scores on a test song.}}
\vspace{-0.3cm}
\label{fig:align}
\end{figure}
\vspace{-0.05cm}
\subsection{Objective evaluation}
Objective tests were conducted to evaluate different SVS systems.  For the convenience of comparison, all systems remained the same ground-truth duration as the target natural audio.
Mel-spectral distortion (MSD), root-mean-square error (RMSE) and correlation coefficients of F0 values on a linear scale between the natural audio and synthesized voice by different SVS systems are presented in Table \ref{tab:obj}. It is worth mentioning that the compared acoustic features  were re-extracted from the generated waveforms. 
The objective result presents that the synthesized voice from the \emph{\textbf{ByteSing}} system can acquire  smallest spectral distortion and achieve more precise pitch prediction. 
Although the tone or intonation modeling in Chinese TTS is quite indispensable, the
comparison between the  \emph{\textbf{ByteSing}} and  \emph{\textbf{BS-w-$\textbf{To}$}} systems shows that the additional tone information can  even  reduce the prediction accuracy of acoustic features. This phenomena may be due to the limited amount of training data and too rich feature representions can also weaken the model generalization ability. Moreover, some acoustic features such as pitch contours  are mainly controlled by the musical notes, which differs with TTS tasks.
The MSD and F0 RMSE of the \emph{\textbf{BS-w/o-atten}} system  are the largest among all the systems. The superiority of the \emph{\textbf{ByteSing}}  system over the \emph{\textbf{BS-w/o-atten}} system  indicates  the importance of using the attention mechanism as the soft alignment manner. Figure \ref{fig:align} illustrates the alignment scores in the attention module along the inference steps. Because the input sequences
are expanded by the auxiliary duration model, the alignment monotonicity is  much more robust than standard end-to-end models and there are nearly no attention errors such as missing and repeating in  the test  phase.

\begin{figure}[t]
\setlength{\belowcaptionskip}{-0.2cm}
\centering
\includegraphics[width=\linewidth]{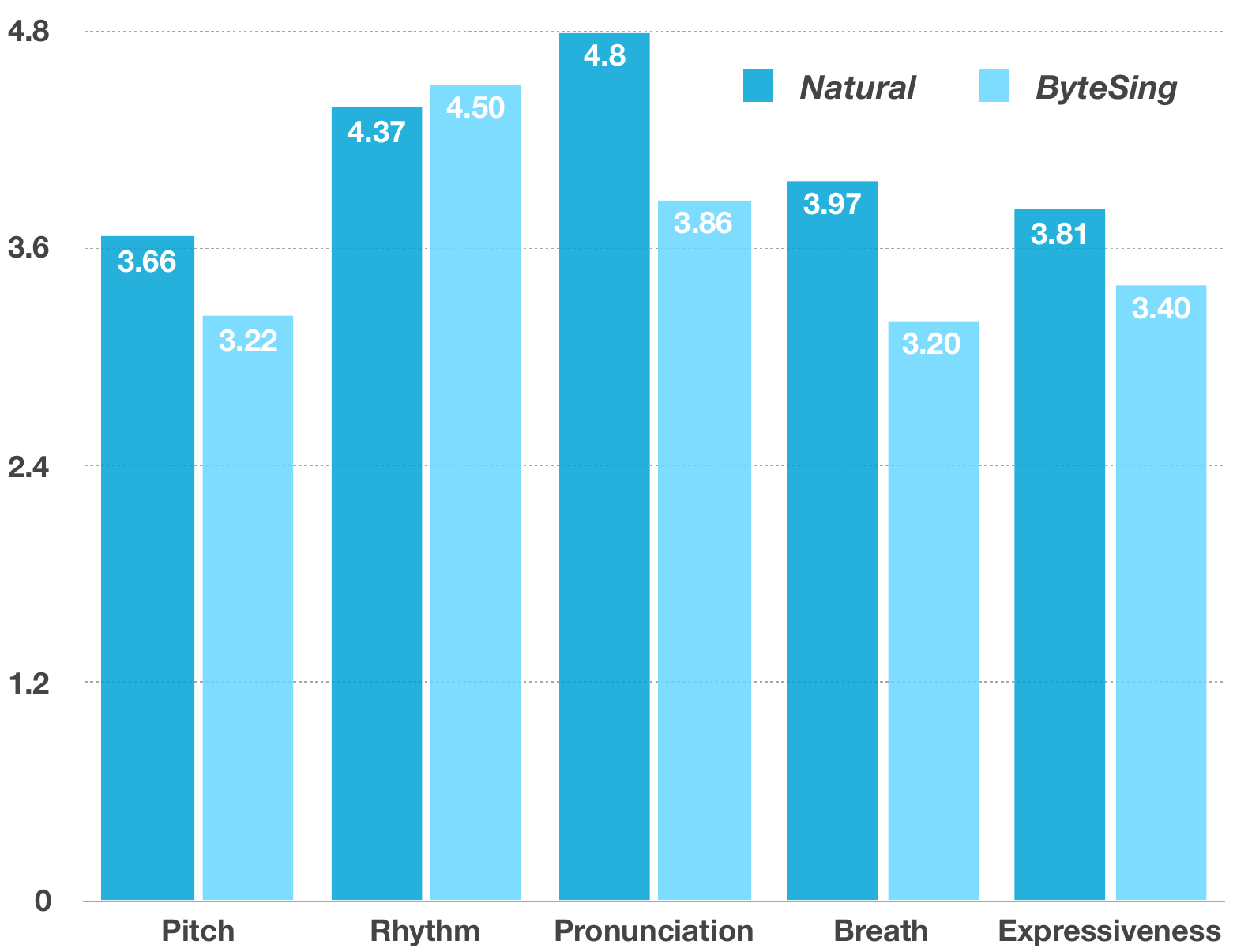}
\vspace{-0.3cm}
\caption{MOSs for both  \textbf{Natural} and  \textbf{ByteSing}  systems.}
\vspace{-0.3cm}
\label{fig:mos}
\end{figure}

\subsection{Subjective evaluation}
To better compare the difference between the synthesized songs and recorded ground-truth songs, five professional  musical experts were employed to evaluate the singing performance of the real singer and ByteSing system respectively. The same assessment standards for vocalists
such as rhythm accuracy, pitch accuracy, pronunciation, breath  and expressiveness are conducted on our synthesized singing voices and mean opinion score (MOS) from 1 (bad) to 5 (excellent) is utilized as a measure for the singing performances.
All the generated songs were synthesized according to the  predicted phoneme duration\footnote{Examples of synthesized speech by different systems are available at \url{https://ByteSings.github.io}.} 

The results of subjective tests are exhibited in  Figure \ref{fig:mos}  respectively.
The MOS comparisons of  the  \emph{\textbf{Natural}} and   \emph{\textbf{ByteSing}} systems  demonstrate the proposed ByteSing can successfully achieve close quality with natural songs. All the measure criteria for ByteSing system  have exceeded 80 percent of these for original recorded songs and the synthesized voices generally  sound quite natural, which proves the effectiveness of the proposed SVS system. The score of intonation or pitch accuracy for the recorded songs is the lowest among all scores. We found that the singer  didn't  always sing at the correct pitch  following the given musical notes and falsetto problems were also serious when singing very high notes, which also decreases the pitch modeling accuracy of our SVS system.  The differences for pronunciation and breath  between   \emph{\textbf{Natural}}  and  \emph{\textbf{ByteSing}} are much larger and more obvious than those of other criteria. We expect to
address these issues by increasing training data volume. The rhythm accuracy is even better than the real singer because of the post-processing procedure on duration prediction, while this can lead in the deficiency for singing expressiveness.



\section{Conclusion}
\label{sec:conclusion}
This paper introduces the proposed ByteSing  system, which adopts Tacotron-like acoustic models and neural vocoders. Mel-spectrograms are directly predicted utilizing encoder-decoder structures and attention modules.  WaveRNNs are employed as vocoders to synthesize waveform directly to exceed the limitation of traditional vocoders. A duration model is also employed to 
improve the robustness, accuracy and  controllability. Subject tests illustrate that ByteSing can achieve more than 80 percent of human singing level.
ByteSing is our first attempt on the
 singing synthesis task. In future work, more training and optimization strategies such as  multiple singer pre-training and data augmentation will be explored  on our ByteSing systems. And a more automatic process which includes auto-labeling, pitch correction and some multi-modal methods will also be developed for the SVS online services.
 

\bibliographystyle{IEEEtran}

\bibliography{mybib}

\end{document}